\documentclass[aps,prl,twocolumn,superscriptaddress]{revtex4-2}

\usepackage[hidelinks]{hyperref}

\usepackage{physics}
\usepackage{balance}
\usepackage{suffix}
\usepackage{mathtools}
\usepackage[utf8]{inputenc}
\usepackage{booktabs}
\usepackage{cases}
\usepackage[multiple]{footmisc}
\usepackage{dcolumn}
\usepackage{color,soul}
\usepackage{rotating}
\usepackage{perpage}
\usepackage{siunitx}
\usepackage{xcolor}
\definecolor{darkcyan}{RGB}{0,139,139}
\hypersetup{
	colorlinks,
	linkcolor={red},
	citecolor={blue},
	urlcolor={blue}
}
\usepackage{soul}
\usepackage{amsmath}
\usepackage{tikz}
\usepackage[T1]{fontenc}
\usepackage{etoolbox}
\usepackage{graphics}
\usepackage{siunitx}
\usepackage{float}	
\usepackage{collref}
\usepackage{multirow}
\usepackage{mathtools}
\usepackage{bm}
\usepackage{url}
\usepackage{pifont}
\usepackage{balance}

\usepackage{amssymb}
\usepackage{mathtools}
\usepackage{amsmath}
\usepackage{bm}
\usepackage{siunitx}
\usepackage{graphicx}
\usepackage{float}
\usepackage{multirow}
\usepackage{booktabs}
\usepackage{tikz}
\usepackage{tikz-3dplot}
\usepackage{pgfplots}
\pgfplotsset{compat=1.18}

\usepackage{tikz}
\usepackage{tikz-3dplot}
\usepackage{accents}

\makeatletter
\newcommand{\doublewidetilde}[1]{{%
		\mathpalette\double@widetilde{#1}%
}}
\newcommand{\double@widetilde}[2]{%
	\sbox\z@{$\m@th#1\widetilde{#2}$}%
	\ht\z@=.9\ht\z@
	\widetilde{\box\z@}%
}
\makeatother

\usepackage{etoolbox,lipsum}

\newcommand{\blue}[1]{\textcolor{black}{#1}}

\begin{document} 
	\title{Floquet-induced anisotropic magnetoresistance and anomalous Hall effect\\ in 2D $d$-wave altermagnets with Rashba spin-orbit coupling}
	\author{Mohsen Yarmohammadi}
	\email{mohsen.yarmohammadi@georgetown.edu}
	\address{Department of Physics, Georgetown University, Washington DC 20057, USA}
	\author{Pieter M. Gunnink}
	\email{pgunnink@uni-mainz.de}
	\affiliation{Institute of Physics, Johannes Gutenberg-University Mainz, Staudingerweg 7, Mainz 55128, Germany}
	\author{Jairo Sinova}
	\email{sinova@uni-mainz.de}
	\affiliation{Institute of Physics, Johannes Gutenberg-University Mainz, Staudingerweg 7, Mainz 55128, Germany}
	\affiliation{Department of Physics, Texas A$\&$M University, College Station, Texas 77843-4242, USA}
	\author{James K. Freericks}
	\email{James.Freericks@georgetown.edu}
	\address{Department of Physics, Georgetown University, Washington DC 20057, USA}
	\date{\today}
	\begin{abstract}
		Altermagnets~(AMs) combine momentum-dependent spin splitting with zero net magnetization, making them promising for spintronics. Periodic driving enables dynamic symmetry engineering beyond static, material-specific control. We show that Floquet engineering in 2D $d$-wave AMs with out-of-plane Néel order and Rashba spin–orbit coupling unlocks equilibrium-forbidden transport responses. Monochromatic driving produces purely out-of-plane magnetization, yielding longitudinal anisotropic magnetoresistance~(AMR) and an anomalous Hall effect, whereas bichromatic driving generates both in-plane and out-of-plane magnetizations and additionally activates transverse AMR via the second harmonic of the secondary beam. Comparable static magnetic fields would require hundreds of tesla, avoided in Floquet driving. These effects persist across linear, circular, and mixed light polarizations and are tunable via light parameters. Our results establish multi-color Floquet engineering for controlling magnetization and symmetry-protected transport in AMs.
	\end{abstract}
	
	\maketitle
	{\allowdisplaybreaks

		\blue{\textit{Introduction}}---Altermagnetism has recently emerged as a magnetic phase distinct from conventional ferromagnets and antiferromagnets~\cite{PhysRevX.12.031042,PhysRevX.12.040501,PhysRevX.12.040002,doi:10.7566/JPSJ.88.123702,PhysRevX.12.011028}. Despite zero net magnetization, altermagnets~(AMs) host symmetry-protected momentum-dependent spin splitting from crystal symmetry and magnetic order~\cite{Krempasky2024,Reimers2024}, with spin polarization forming $d$-, $g$-, or $i$-wave harmonics. In particular, $d$-wave systems exhibit a strong spin-splitter effect and enable spin-polarized transport without stray fields, with tunable control via electrical and strain fields~\cite{Bose2022,PhysRevLett.128.197202,PhysRevLett.129.137201,PhysRevLett.126.127701,doi:10.1126/sciadv.adn0479,PhysRevB.109.144421,Song2025,jungwirth2025altermagneticspintronics,Jiang2025,PhysRevLett.132.176702,Naka2019,PhysRevLett.126.127701,zarzuela2024transporttheoryspintransferphysics}. Spintronics broadly aims to generate and control spin-polarized currents~\cite{RevModPhys.76.323}, typically through spin pumping or charge-to-spin conversion mediated by spin--orbit coupling~(SOC)~\cite{RevModPhys.77.1375,RevModPhys.87.1213,10.1063/1.3587173}. However, many conventional spintronic mechanisms rely on relativistic SOC and magnetic dynamics, restricting their operation to GHz–THz timescales. AMs instead offer symmetry-controlled spin transport without net magnetization or stray fields, facilitating a route to ultrafast control.
		
		Interfacial Rashba SOC~(RSOC) and AMs provide a platform for transport, where the resulting in-plane helical spin texture combined with altermagnetic spin splitting enables spin-valve and filtering effects without net magnetization~\cite{jqg2-fpk3,darvishi2026exploringconventionalanomalousjosephson,mizoguchi2026orbitalzeemancrosscorrelationp,fakhredine2026interplayrelativisticspinmomentumlocking,yarmohammadi2026efficienttwocolorfloquetcontrol,acharjee2026unconventionalspinvalveeffect,MARFOUA202547,PhysRevB.111.174431,pzt5-5ypz}. In 2D $d$-wave AMs with RSOC~\cite{sheoran2025tuningspincurrentscollinear,3sw6-y8vf,342f-82rj,w52v-blqm,zkfz-zh4l}, the underlying tetragonal symmetries of the 2D plane strongly constrain transport. Symmetry reduction via strain~\cite{r7mn-yklk,Li2025,PhysRevB.109.144421,8zlt-mlms,zhang2026nearlycompletechargespinconversion}, substrate/interface engineering~\cite{hoi2026rkkyinteractionaltermagnetsadiabatic,4txy-gy8f,zhang2026keyrolechargedisproportionation,choi2026orbitaldrivenemergenttransportaltermagnets,liu2026altermagnetismroomtemperaturemetaltoinsulatortransition,zhao2026layerdependentquantumtransportkv2se2obased,Jackel2026femtosecondallopticalcoherentcontrol,sharma2026doublepeakmajoranaboundstates}, disorder~\cite{darvishi2026exploringconventionalanomalousjosephson,woodgate2026lossaltermagneticordersmooth,varshney2026asymmetricscatteringdriveslarge}, or Zeeman fields~\cite{shaffer2025theoriessuperconductingdiodeeffects,kundu2025rkkyinteractionmediatedspinpolarized,ZHANG2026100379} lift these constraints and induce anisotropic or transverse responses, but remain static and material-specific. Achieving strong control typically requires magnetic fields of hundreds of tesla, beyond dc limits. In comparison, optical driving enables reversible symmetry engineering and tunable transport beyond equilibrium constraints~\cite{doi:10.7566/JPSJ.94.111007,annurev-conmatphys-031218-013423,PhysRevResearch.4.033213,Rudner2020,Giovannini_2020,yu2026tunableoddparityspinsplittings,yarmohammadi2025anisotropiclighttailoredrkkyinteraction,yarmohammadi2026efficienttwocolorfloquetcontrol,gill2026opticalultrafastpurespin,chen2026routenonrelativisticaltermagneticspin,tian2026opticallydrivenorbitalhall,liu2026linearlypolarizedlightinducedanomalous,wnqs-3djt,yang2025nonlinearoptomagneticsignaturedwave,qin2026anomalousthermoelectricthermalhall,xzm1-l6yf} and static methods.
		
		Despite recent advances in Floquet engineering of AMs~\cite{yarmohammadi2025anisotropiclighttailoredrkkyinteraction,yarmohammadi2026efficienttwocolorfloquetcontrol,gill2026opticalultrafastpurespin,chen2026routenonrelativisticaltermagneticspin,tian2026opticallydrivenorbitalhall,liu2026linearlypolarizedlightinducedanomalous,wnqs-3djt,yang2025nonlinearoptomagneticsignaturedwave,qin2026anomalousthermoelectricthermalhall,tian2026opticallydrivenorbitalhall,xzm1-l6yf}, it is still not understood how \textit{multi-frequency} driving and harmonic interference break symmetry constraints to produce transverse charge and spin transport that is otherwise forbidden.\begin{figure}[t]
			\centering
			\includegraphics[width=1\linewidth]{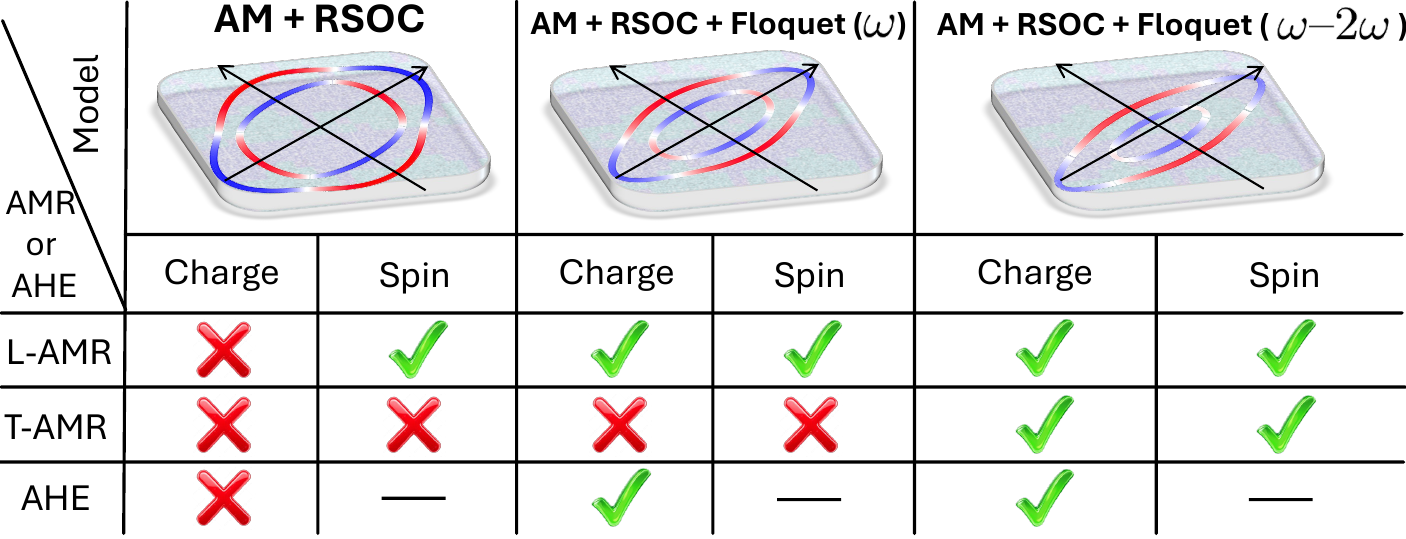}
			\caption{Schematic of AMR and AHE in a 2D $d_{x^2-y^2}$-wave AM with an out-of-plane Néel vector and RSOC under Floquet driving. Red and blue contours denote spin-split Fermi surfaces. In equilibrium (AM + RSOC), the system has $C_{2v}$ symmetry, which forbids charge L-AMR, T-AMR, and AHE, while allowing spin L-AMR and enforcing vanishing spin T-AMR. Under monochromatic driving ($\omega$), the symmetry is reduced but remains $C_{2v}$-compatible, allowing charge L-AMR and AHE while T-AMR is still forbidden. In contrast, bichromatic driving ($\omega$--$2\omega$) further reduces the symmetry from $C_{2v}$ toward $C_1$, thereby activating both charge and spin T-AMR.}
			\label{f1}
		\end{figure}
		
		In this Letter, we combine Floquet--Magnus theory, Boltzmann transport, and the Kubo–Berry curvature framework in 2D AMs with an out-of-plane Néel vector and RSOC under multi-color driving, to lift symmetry constraints on charge and spin transport. In equilibrium, $C_{2v}$ symmetry forbids charge longitudinal anisotropic magnetoresistance~(L-AMR), transverse AMR~(T-AMR), and anomalous Hall effect~(AHE), while allowing spin L-AMR and vanishing spin T-AMR~[Fig.~\ref{f1}, left column]. Under monochromatic driving~($\omega$), $C_{2v}$ is preserved, enabling charge L-AMR and AHE but keeping T-AMR forbidden [Fig.~\ref{f1}, middle column]. However, bichromatic~($\omega$--$2\omega$) driving reduces the symmetry toward $C_1$ via photon-sector interference, activating both charge and spin T-AMR [Fig.~\ref{f1}, right column].
		
		\blue{\textit{Hamiltonian model}}---We consider a 2D $d$-wave AM with RSOC strength $\lambda$, described near the $\Gamma$ point by
		\begin{equation}
			\label{eq:H0_main}
			H_d(\mathbf{k})=
			\varepsilon k^2\sigma_0
			+\lambda(k_x\sigma_y-k_y\sigma_x)
			+ \varepsilon M_{\mathbf{k}}^d\sigma_z ,
		\end{equation}
		where $\varepsilon=\hbar^2/(2m_e)$ and\begin{equation}\label{eq_2}
			M_{\mathbf{k}}^d=
			M_{1g}(k_x^2-k_y^2)
			+2M_{2g} k_xk_y\,.
		\end{equation}Here $M_{1g}=M_d\cos(2\theta)$ and $M_{2g}=M_d\sin(2\theta)$, with $M_d$ and $\theta$ denoting the strength and orientation of the altermagnetic order parameter. For tetragonal symmetry, $(k_x^2-k_y^2)$ and $2k_xk_y$ transform as $B_{1g}$ and $B_{2g}$ of $D_{4h}$ point group, making Eq.~\eqref{eq_2} a general $d$-wave altermagnetic order combining the two quadrupolar channels, with the corresponding spin point group $^2 4/m$.
		
		To generate dynamical Floquet states, we couple the system to a bichromatic light via the Peierls substitution
		$\mathbf{k}\rightarrow \mathbf{k}+e\mathbf{A}(t)/\hbar$ with a general time-periodic vector potential{\small\begin{subequations}
			\begin{align}
				A_x(t) &= \mathcal{A}\!\left[\cos(\omega\,t)
				+ \mathcal{S}\cos(n\omega\,t+\phi)\cos\eta\right],\\
				A_y(t) &= \mathcal{A}\!\big[c\sin(\omega\,t)
				+ \tilde{c}\,\mathcal{S}\cos{\small(}n\omega\,t+\phi - \tilde{\phi}{\small)}\cos(\eta - \tilde{\eta})\big],
			\end{align}
		\end{subequations}}where $\mathcal{A}$ sets the field strength, $\omega$ is the fundamental frequency, $n$ is the harmonic ratio, $\mathcal{S}$ is the relative amplitude, and $\{\phi,\tilde{\phi}\}$ are the inter-color phases. The parameters $\{\eta,\tilde{\eta}\}$ control the light polarization angles, while $\{c,\tilde{c}\}$ encode the light polarization character of each frequency component. For the bilinear case, we set $\tilde{\phi}=0$, $c=0$, $\tilde{c}=1$, $\eta\neq0$, and $\tilde{\eta}=\pi/2$; for the bicircular case, we choose $\tilde{\phi}=\pi/2$, $\{c,\tilde{c}\}=\pm1$, $\eta=0$, and $\tilde{\eta}=0$; and for the hybrid case, we take $\tilde{\phi}=0$, $c=\pm1$, $\tilde{c}=1$, $\eta\neq0$, and $\tilde{\eta}=\pi/2$. This driving enables tunable band topology and symmetry breaking~\cite{doi:10.7566/JPSJ.94.111007,annurev-conmatphys-031218-013423,PhysRevResearch.4.033213,Rudner2020,Giovannini_2020}. 
		
		In the off-resonant regime ($\hbar\omega$ exceeds all intrinsic energy scales), real absorption is suppressed, and within the idealized limit of our model, the dynamics is governed by virtual-photon processes. The stroboscopic evolution is described by a high-frequency Floquet-Magnus expansion, $H_{\rm eff}=H_0^{\rm F}+\sum_{m\ge1}\frac{1}{m\hbar\omega}[H_{-m}^{\rm F},H_{+m}^{\rm F}]$, where $H_m^{\rm F}$ are Fourier components of $H_d(\mathbf{k}+e\mathbf{A}(t)/\hbar)$. For the two-band spinful system, Floquet corrections arise from spin-sector commutators, reflecting interference between spin textures mediated by virtual photons. Since the $d$-wave altermagnetic order is quadratic in momentum, the drive generates harmonics up to $m=\pm2$, with the second harmonic ($n=2$) providing the dominant coupling.
		
		The resulting Floquet-engineered system is then described by an effective Hamiltonian of the form $H_{\rm eff}(\mathbf{k}) = h_0(\mathbf{k})\sigma_0 + \mathbf{h}(\mathbf{k})\cdot\boldsymbol{\sigma}$. For the three driving configurations~(bilinear, bicircular, or hybrid circular-linear), the coefficients take the following forms:\begin{subequations}\label{eq_4}
			\begin{align}
				h_0 &= \varepsilon (k_x^2+k_y^2) + \delta h_0^{\rm pol},\label{eq_4a}\\
				h_x &= -\lambda k_y + \delta h_x^{\rm pol}  + \delta h_x^{\rm pol}(\textbf{k}),\label{eq_4b}\\
				h_y &= \lambda k_x + \delta h_y^{\rm pol} + \delta h_y^{\rm pol}(\textbf{k}),\label{eq_4c}\\
				h_z &= \varepsilon M_{\mathbf{k}}^d + \delta h_z^{\rm pol},\label{eq_4d}
			\end{align}
		\end{subequations}where $\delta h^{\rm pol}$ and $ \delta h^{\rm pol}(\textbf{k})$~[see Eqs.~\eqref{eq_13}--\eqref{eq_15} in the End Matter and Supplemental Materials~(SM)~\cite{SM} for the details] denote Floquet-induced corrections depending on the drive polarization; each light beam selects a $d$-wave pattern, either pure (e.g., $B_{1g}$ or $B_{2g}$) or a superposition thereof. They encode three physically distinct effects: (i) AC Stark renormalization of the band dispersion~[Eq.~\eqref{eq_4a}], (ii) modification of RSOC~[last terms of Eqs.~\eqref{eq_4b} and~\eqref{eq_4c}], and (iii) emergent Zeeman-like fields, generating magnetizations $\mathcal{M}_{x,y,z}$~[second terms of Eqs.~\eqref{eq_4b}--\eqref{eq_4d}] arising from interference between RSOC and the $d$-wave altermagnetic order. The $\delta h_x^{\rm pol}$ and $\delta h_y^{\rm pol}$ terms arise solely from the second beam, whereas both beams share all other corrections. These terms vanish in the absence of either altermagnetism or RSOC, and are therefore unique to the Floquet-driven AMs with RSOC.\begin{figure}[t]
			\centering
			\includegraphics[width=0.85\linewidth]{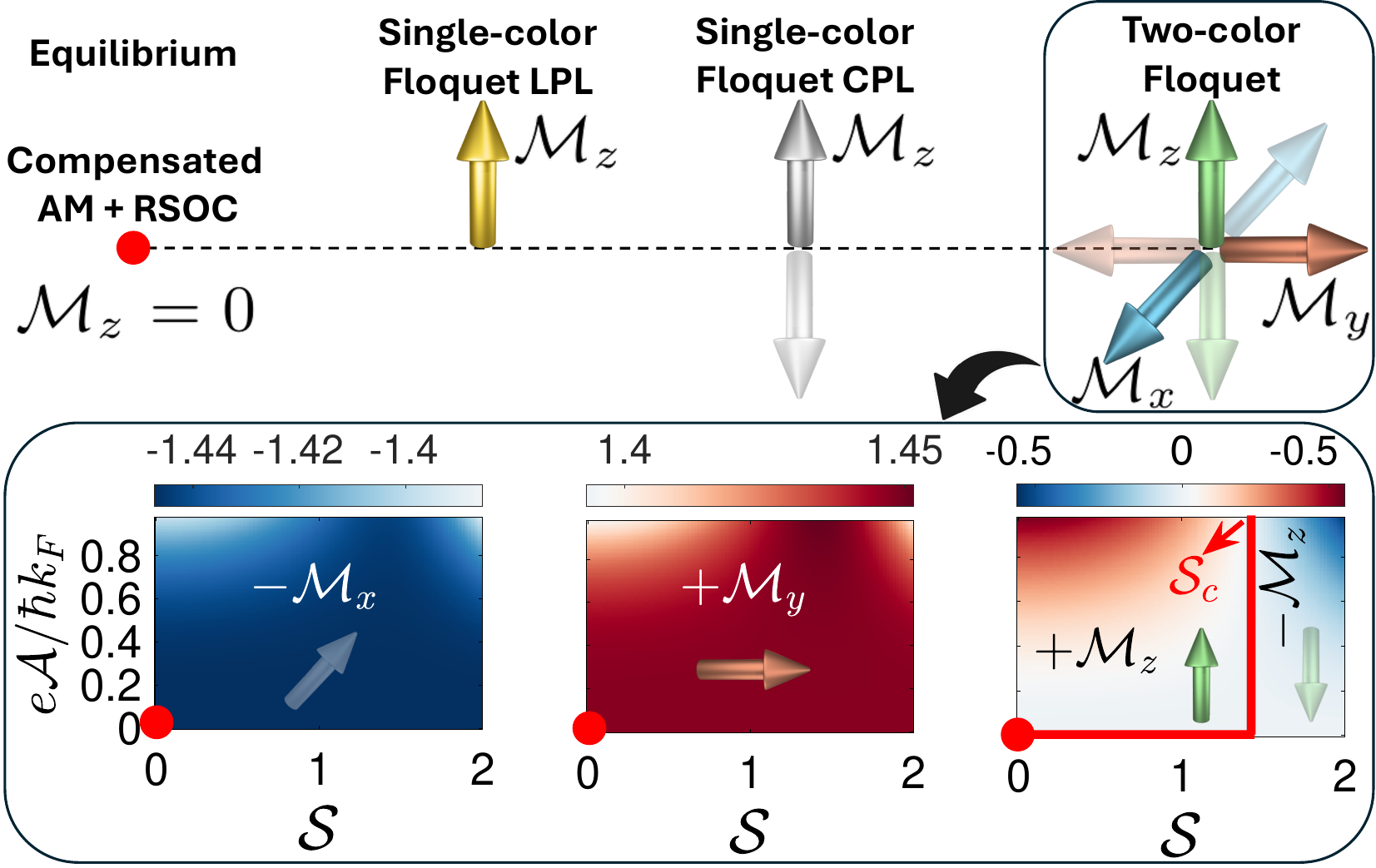}
			\caption{(Top) Schematic of Floquet-induced magnetization in a compensated 2D $d_{x^2-y^2}$-wave AM with an out-of-plane Néel vector and RSOC. Equilibrium has zero net magnetization due to the compensation of spin-split sectors. Linearly polarized light~(LPL) induces finite $\mathcal{M}_z$, while circularly polarized light~(CPL) generates a chirality-dependent~($c = \pm1$) $\mathcal{M}_z$; bichromatic driving enables full vector magnetization. (Bottom) Phase diagram of two-color LPL-induced magnetizations~[color bar in units of $0.01 \mu_{\rm B}$] for $\eta=\pi/3$, $M_{1g}=1$, $M_{2g}=0$. The red dot and line mark the compensated state with zero net magnetization. $\mathcal{M}_x$ and $\mathcal{M}_y$ differ roughly by a sign, while $\mathcal{M}_z$ vanishes at $e\mathcal{A}/\hbar k_F=0$ up to $\mathcal{S}_c=\sqrt{2}$ and changes sign for $\mathcal{S}>\mathcal{S}_c$.}
			\label{f2}
		\end{figure} 
		
		Figure~\ref{f2}(top) summarizes the magnetization response of a compensated AM with an out-of-plane N\'eel vector and RSOC. The magnetization is computed from the occupied Floquet states with eigenvalues $\epsilon_\nu(\mathbf{k})$ and eigenvectors $|u_\nu(\mathbf{k})\rangle$ of band $\nu$,
		$\mathcal{M}_{x,y,z}=\sum_{\nu}\!\int\!\frac{d^2k}{(2\pi)^2}
		f_\nu(\mathbf{k})
		\langle u_{\nu\mathbf{k}}|\sigma_{x,y,z}|u_{\nu\mathbf{k}}\rangle$, where $f_\nu(\mathbf{k})$ denotes the distribution function. In equilibrium, spin-split contributions cancel, yielding zero net magnetization. Single-color linearly polarized light~(LPL) induces a finite out-of-plane magnetization $\mathcal{M}_z$, while circularly polarized light~(CPL) generates a chirality-dependent ($c=\pm1$) $\mathcal{M}_z$. On the other hand, bichromatic driving relaxes the symmetry constraints and enables a full vector magnetization. Figure~\ref{f2}(bottom) shows the phase diagram obtained from Eqs.~\eqref{eq_13b}--\eqref{eq_13d} in the End Matter. The in-plane components satisfy $\mathcal{M}_x \approx -\mathcal{M}_y$, while $\mathcal{M}_z$ vanishes for $e\mathcal{A}/\hbar k_F = 0$ up to $\mathcal{S}_c = \sqrt{2}$; at which $\delta h_z = 0$ for $\cos(2\eta) = -1/2$, and it changes sign for $\mathcal{S} > \mathcal{S}_c$, indicating a symmetry-controlled Floquet transition.
		
		To relate the dimensionless drive, $e\mathcal{A}/\hbar k_F \sim 0.1$--$0.9$ in our simulations, to experiments, we estimate field strengths using $k_F=\sqrt{2m_e|\mu_0|}/\hbar$ with $\mu_0\simeq 0.5$ eV ($k_F\approx 1~\text{\AA}^{-1}$). For a drive frequency $\hbar\omega\simeq 3$ eV, the electric field $E_0=\omega \mathcal{A}$ scales as $E_0\propto \omega\sqrt{|\mu|}$, yielding $E_0\sim 1.5$--$3~\mathrm{V/nm}$. These values lie within experimentally accessible ultrafast pump regimes where Floquet and nonlinear effects are prominent. The effective Zeeman-like fields reach $\sim 10$--$30$ meV for the above realistic drive parameters, corresponding to magnetic fields of order $\sim 100$ T, far beyond experimentally accessible dc limits. This highlights the key advantage of Floquet engineering: the generation of large, tunable, symmetry-selective effective spin textures inaccessible in equilibrium without huge static magnetic fields.
		
		\blue{\textit{Charge and spin conductivity}}---Next, we study charge and spin transport in a system described by a general Bloch Hamiltonian $H(\mathbf{k})$. Transport is computed within the Boltzmann formalism in the constant relaxation-time approximation $\tau$~\cite{RevModPhys.87.1213,RevModPhys.82.1959,RevModPhys.82.1539,ashcroft1976solid,ziman1960electrons}. Semiclassical wave packets under an external electric field $\mathbf{E}$ evolve as $\dot{\mathbf{r}} = \frac{1}{\hbar}\nabla_{\mathbf{k}} \epsilon_\nu(\mathbf{k})$ and $\dot{\mathbf{k}} = -\frac{e}{\hbar}\mathbf{E},$
		with group velocity $\mathbf{v}_\nu(\mathbf{k})=\hbar^{-1}\nabla_{\mathbf{k}}\epsilon_\nu(\mathbf{k})$. The charge~(c) and spin~(c), for $s_z = \hbar \sigma_z/2$, current is $\mathbf{j}^c = -e \sum_\nu \int \frac{d^2 k}{4\pi^2}
		\, \mathbf{v}_\nu(\mathbf{k}) f_\nu(\mathbf{k})$ and $\mathbf{j}^s = \sum_\nu \int \frac{d^2 k}{4\pi^2}
		\, \mathbf{v}_\nu(\mathbf{k})\, s_\nu(\mathbf{k})\, f_\nu(\mathbf{k})$, respectively, where $s_\nu(\mathbf{k})=\langle u_{\nu\mathbf{k}}|\hat{s}_z|u_{\nu\mathbf{k}}\rangle$. This yields the Fermi-surface conductivity tensors~($\ell , j \in \{x,y\}$) for AMR at zero temperature{\small\begin{subequations}
				\begin{align}
					\sigma_{\ell j}^c = {} &e^2 \tau \sum_\nu \int \frac{d^2 k}{4\pi^2}
					\, v_{\nu,i} v_{\nu,j}
					\, \delta(\epsilon_\nu(\mathbf{k}) - \mu),\\
					\sigma_{\ell j}^s = {} &e \tau \sum_\nu \int \frac{d^2 k}{4\pi^2}
					\, v_{\nu,i} v_{\nu,j} s_\nu(\mathbf{k})
					\, \delta(\epsilon_\nu(\mathbf{k}) - \mu)\, .
				\end{align}
		\end{subequations}}The L-AMR quantifies anisotropy along the principal axes, $\mathrm{L\text{-}AMR}=
		(\sigma_{xx}-\sigma_{yy})/(\sigma_{xx}+\sigma_{yy})$, capturing the imbalance of \(v_x^2\) and \(v_y^2\) Fermi-surface contributions. The T-AMR characterizes the off-diagonal symmetric response, $\mathrm{T\text{-}AMR}=
		2\sigma_{xy}/(\sigma_{xx}+\sigma_{yy})$, arising from \(\langle v_x v_y \rangle\) correlations. Here \(\sigma_{xy}^{c/s}=\sigma_{yx}^{c/s}\), distinguishing this contribution from the Berry-curvature-induced AHE.
		
		The low-temperature AHE is obtained from the Berry curvature, $\Omega_\nu(\mathbf{k})$, of the Floquet-dressed bands as\begin{align}
			\sigma_{xy}^{\mathrm{AHE}} = -\frac{e^2}{\hbar} \sum_\nu \int \frac{d^2k}{4\pi^2} \Theta(\mu - \epsilon_\nu(\mathbf{k})) \Omega_\nu(\mathbf{k}),
		\end{align}where $\Theta(\mu-\epsilon)$ is the Heaviside step function.
		
		\blue{\textit{Results and discussion}}---We now present numerical results using Gaussian broadening of $\delta(\epsilon-\mu)$ and a smearing which corresponds to a temperature of 0.5 meV. For AMR, all charge (spin) conductivities are normalized by $e^2\tau/\pi\hbar^2$ ($e\tau/\pi\hbar$), while the AHE is reported in units of $e^2/h$. To avoid divergence issues in the spin sector, we normalize using $|\sigma_{xx}^{s}|+|\sigma_{yy}^{s}|$, defining spin  $\mathrm{L\text{-}AMR}=2(\sigma^{s}_{xx}-\sigma^{s}_{yy})/(|\sigma^{s}_{xx}|+|\sigma^{s}_{yy}|)$ and spin $\mathrm{T\text{-}AMR}=2\sigma^{s}_{xy}/(|\sigma^{s}_{xx}|+|\sigma^{s}_{yy}|)$. Moreover, Floquet corrections show bichromatic interference governed by $\eta$ and $\phi$, which we fix at $\eta=\pi/3$ and $\phi=\pi/6$ without qualitatively affecting the results.
		
		The point-group symmetry of the effective Floquet Hamiltonian governs the transport properties. In the absence of altermagnetic order, the Rashba system possesses $C_{4v}$ symmetry (or $D_{4h}$ for a centrosymmetric parent lattice). The $d$-wave exchange field $M_{\mathbf{k}}^d$, transforming as the $B_{1g}$ representation, breaks the fourfold rotational symmetry and reduces the point group to $C_{2v}$, removing the $x\leftrightarrow y$ equivalence. The residual symmetry $C_{2v}=\{C_{2z},m_x,m_y\}$ consists of a twofold rotation about the $z$ axis, $C_{2z}$, and mirror reflections $m_x:(k_x,k_y)\!\to\!(k_x,-k_y)$ and $m_y:(k_x,k_y)\!\to\!(-k_x,k_y)$. Single-color driving preserves this residual symmetry, whereas bichromatic $\omega$--$2\omega$ driving generates the in-plane Floquet fields $\delta h_x^{\rm pol}$ and $\delta h_y^{\rm pol}$ [Eq.~\eqref{eq_4}], reducing the symmetry to the trivial group $C_1$ (identity only). The resulting symmetry hierarchy, $C_{4v}\rightarrow C_{2v}\rightarrow C_1,$ underlies the emergence of anisotropic longitudinal and transverse transport responses.
		
		The charge and spin conductivities of the pristine 2D $d$-wave AM with RSOC follow from Fermi-surface velocity correlators [Eqs.~\eqref{eq_15n}--\eqref{eq_19n} in the End Matter]. In the absence of RSOC and driving, the charge response is isotropic, $\sigma_{xx}^c=\sigma_{yy}^c=e^2\tau\mu/\pi\hbar^2\sqrt{1-M_d^2}$, yielding zero charge L-AMR, while the $d$-wave texture produces an anisotropic spin response, $\sigma_{xx}^s=-\sigma_{yy}^s=e\tau\mu M_d\cos(2\theta)/4\pi\hbar\sqrt{1-M_d^2}$ with $\theta$ being the orientation of AM, resulting in finite spin L-AMR. The transverse response is purely spin, with $\sigma_{xy}^c=0$ and $\sigma_{xy}^s=-e\tau\mu M_d\sin(2\theta)/4\pi\hbar\sqrt{1-M_d^2}$, leading to zero~(finite) spin T-AMR for $d_{x^2-y^2}$-wave~($d_{xy}$-wave) AM~($\theta = \pi/4$).\begin{figure}[t]
			\centering
			\includegraphics[width=0.85\linewidth]{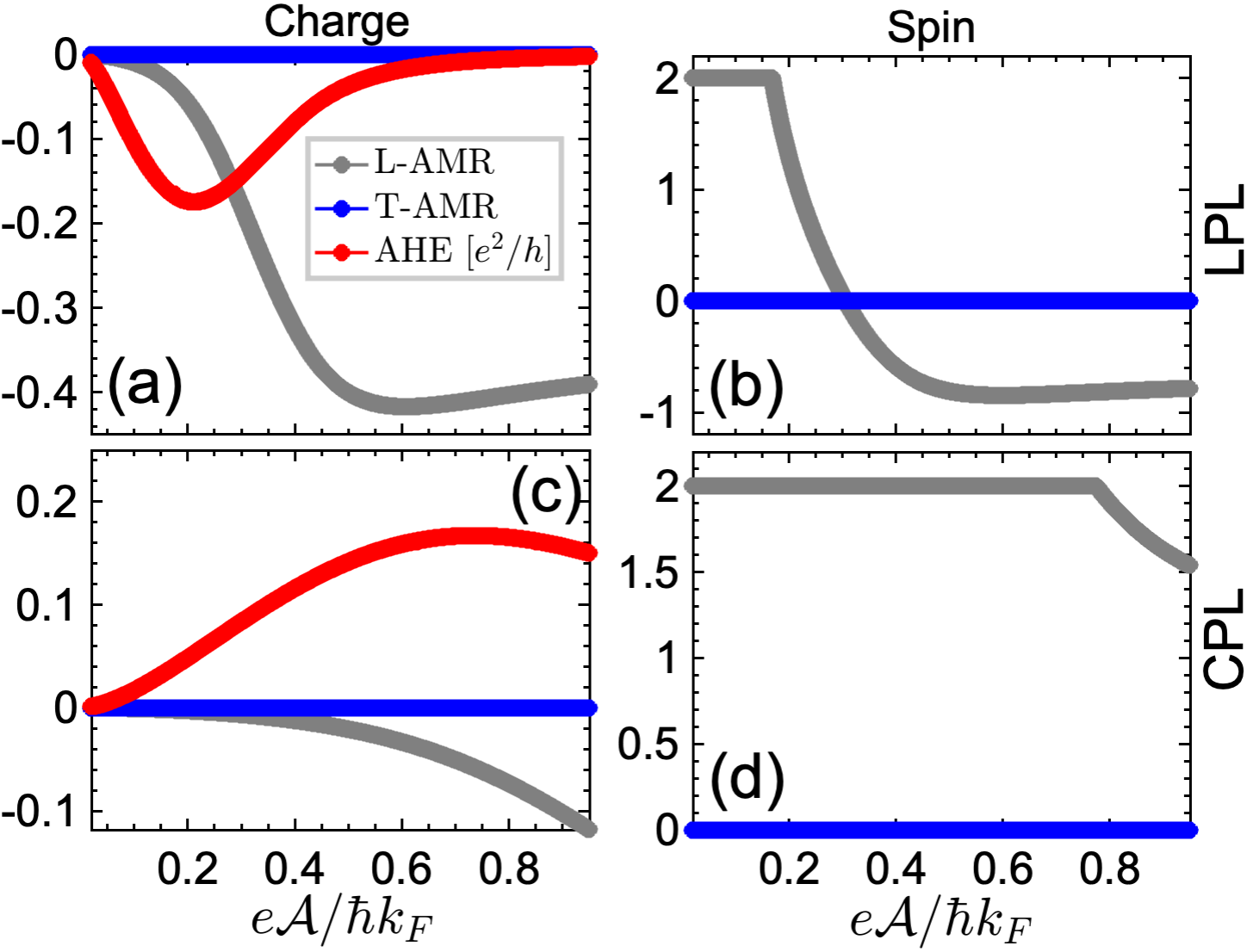}
			\caption{Charge (left column) and spin (right column) AMR responses of a $d_{x^2-y^2}$-wave AM with an out-of-plane Néel vector and RSOC under monochromatic Floquet driving for linear (a,b) and circular (c,d) polarization as a function of $e\mathcal{A}/\hbar k_F$. The system retains $C_{2v}$ symmetry under single-frequency driving without T-AMR, while allowing only L-AMR and AHE. Floquet-induced band dressing generates strong nonlinear anisotropic L-AMR in both charge and spin sectors, with light polarization-controlled modulation and reversal of spin L-AMR. No T-AMR emerges, consistent with symmetry constraints. Parameters are $M_d=0.7$, $\lambda=0.2$ eV\,\AA, $c=+1$ for CPL, and $\mathcal{S}=0$.}
			\label{f3}
		\end{figure}
		
		Figure~\ref{f3}(a--d) isolates single-frequency Floquet driving in a $d_{x^2-y^2}$-wave AM with an out-of-plane Néel vector and RSOC, where the response is constrained by the underlying $C_{2v}$ symmetry. In equilibrium ($e\mathcal{A}/\hbar k_F = 0$), this symmetry enforces the absence of L-AMR, T-AMR, and AHE signals~[Fig.~\ref{f3}(a,b)]. Under monochromatic driving, Floquet-induced band mixing preserves the $C_{2v}$-compatible selection rules, leading to nonlinear enhancement of L-AMR and AHE without activating T-AMR. Both charge and spin sectors exhibit pronounced light polarization-dependent modifications of AMR: LPL reverses the spin L-AMR, whereas CPL produces only a weak modulation of L-AMR at stronger driving amplitudes~[Fig.~\ref{f3}(b,d)]. Nevertheless, T-AMR remains symmetry-forbidden for both light polarizations.
		
		From Eqs.~\eqref{eq_4b}--\eqref{eq_4c}, the second beam generates Floquet-induced in-plane fields $\delta h_x^{\rm pol}$ and $\delta h_y^{\rm pol}$ that break the residual $C_{2v}$ symmetry preserved under monochromatic driving. The effective Floquet Hamiltonian is reduced toward $C_1$ symmetry, lifting constraints that forbid transverse transport and enabling finite charge and spin T-AMR. Since the AHE is already present under monochromatic driving and is only further modified by the second beam~[see Fig.~\ref{f5} in the End Matter], we focus here on T-AMR.
		
		Figure~\ref{f4}(a,b) shows the resulting T-AMR response as a function of the second beam amplitude $\mathcal S$. For bilinearly polarized light~(BLPL), the symmetry reduction is relatively weak and appreciable T-AMR emerges only when the two frequency components become comparable, $\mathcal{S} \approx 1$. At \(S\approx\sqrt{2}\), where $\delta h_z = 0$ in Eq.~\eqref{eq_13d}, the Floquet-renormalized \(\mathbf{h}(\mathbf{k})\) yields maximal anisotropic velocity mixing, enhancing charge \(\sigma_{xy}^c\), while the spin texture \(s_z(\mathbf{k})=h_z(\mathbf{k})/|\mathbf{h}(\mathbf{k})|\) becomes symmetry-compensated, causing cancellation of \(\sigma_{xy}^s\). In contrast, bicircularly polarized light~(BCPL) generates stronger Floquet interference between Rashba and altermagnetic textures, producing a much larger transverse response. The hybrid BCLPL configuration combines linear and chiral Floquet channels, leading to pronounced nonmonotonic behavior and enhanced tunability of both charge and spin T-AMR. Although the induced in-plane fields scale linearly with $\mathcal S$, the transverse response is governed by mixed velocity correlators $\langle v_x v_y \rangle$, yielding an overall nonlinear dependence on the drive strength.\begin{figure}[t]
			\centering
			\includegraphics[width=0.9\linewidth]{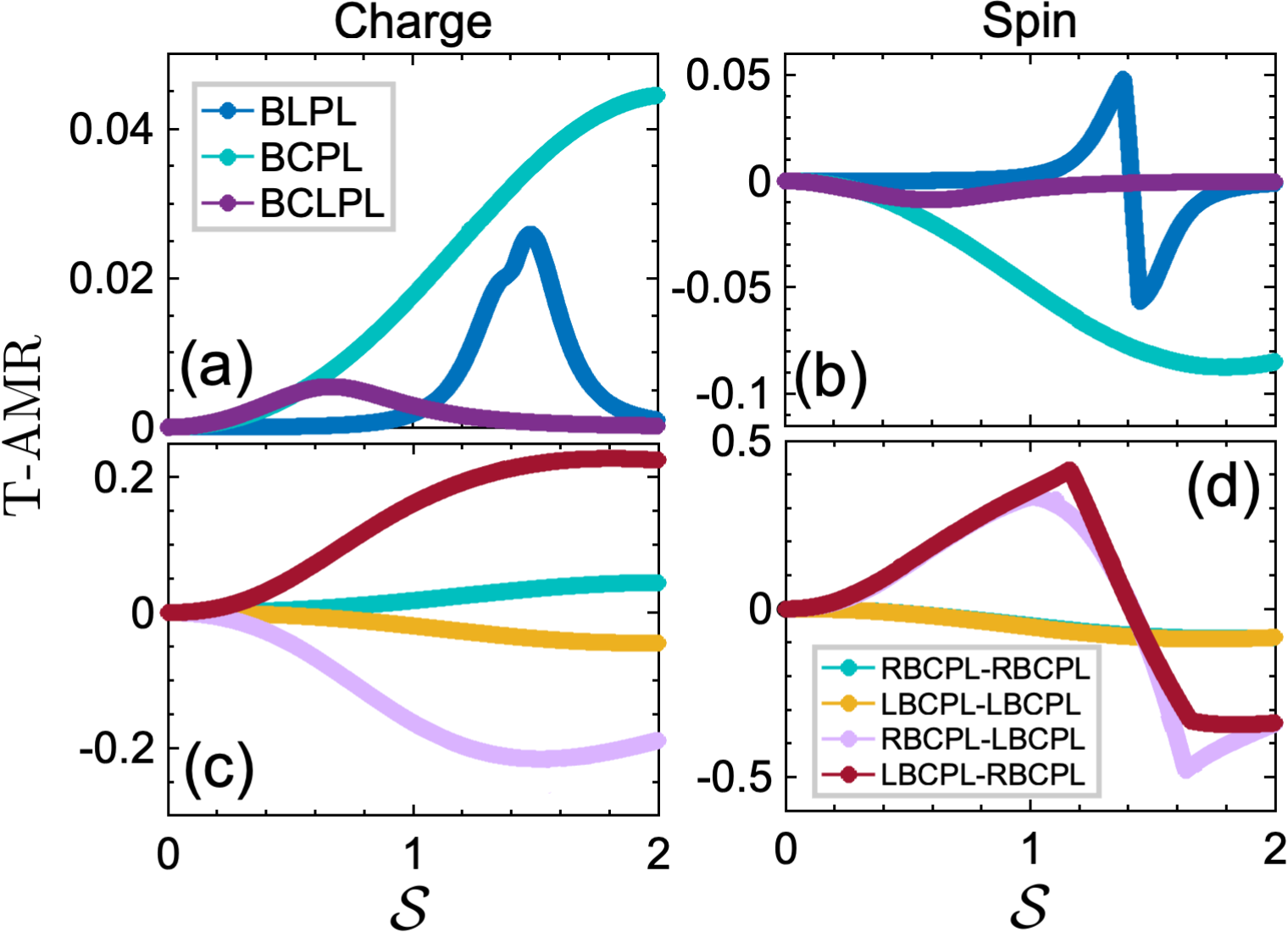}
			\caption{T-AMR responses of a $d_{x^2-y^2}$-wave AM with an out-of-plane Néel vector and RSOC as functions of the second beam amplitude $\mathcal{S}$ at fixed $e\mathcal{A}/\hbar k_F=0.7$. Charge (left column) and spin (right column) T-AMR are shown for BLPL, BCPL with $c=\tilde c=+1$, and hybrid BCLPL driving [(a),(b)] with $c=+1$, and for different circular-field chiralities [(c),(d)]. Unlike monochromatic driving, bichromatic $\omega$--$2\omega$ fields reduce the symmetry from $C_{2v}$ toward $C_1$, activating finite transverse charge and spin responses. The T-AMR depends strongly on light polarization configuration and circular-field chirality. Parameters are $M_d=0.7$, $\lambda=0.2$ eV\,\AA.}
			\label{f4}
		\end{figure}
		
		We next vary the chirality of the circular beams [Figs.~\ref{f4}(c,d)]. The chirality pair $\{c,\tilde c\}$ controls interference terms that determine whether Rashba--altermagnetic mixing is constructive or destructive. Opposite chiralities maximize the symmetry-breaking interference and produce the largest charge T-AMR, while equal chiralities substantially suppress it. Conversely, the spin response is less sensitive to chirality and remains largely chirality-even. Opposite chiralities lead to strong compensation of the Berry curvature and a sign reversal near $\mathcal{S}\approx\sqrt{2}$, where $\delta h_z = 0$ in Eq.~\eqref{eq_14d}. Thus, beyond activating transverse transport through the reduction $C_{2v}\rightarrow C_1$, the chirality of the driving fields provides an efficient handle for controlling the magnitude and sign of Floquet-induced T-AMR.
		
		So far, we have focused on the $d_{x^2-y^2}$-wave AM; other crystalline orientations $\theta$ (e.g., $\theta=\pi/4$ for $d_{xy}$-wave AM) are shown in Fig.~\ref{f6} in the End Matter. It should also be noted that the Floquet-engineered transport phenomena can be readily extended to $g$- and $i$-wave AMs with quartic and sextic momentum structures, leading to distinct $\delta h^{\rm pol}$ and $\delta h^{\rm pol}(\mathbf{k})$ corrections, but the same conclusions.
		
		\blue{\textit{Experimental perspective}}---The predicted Floquet-engineered transport phenomena can be probed in thin-film AMs, such as KV$_2$Se$_2$O~\cite{Jiang2025}, KRu$_4$O$_8$~\cite{PhysRevX.12.031042,Weber2025}, and Mn$_5$Si$_3$~\cite{Reichlova2024}, where interfacial inversion breaking and strain or substrate engineering can realize the required Rashba coupling and tunable $d$-wave order. Experimentally, bichromatic pump--probe transport with phase-locked $(\omega,2\omega)$ visible/near-IR optical pulses, combined with a weak dc bias and lock-in detection, provides direct access to the predicted nonlinear charge and spin responses, building on established techniques for ultrafast dc transport in Floquet systems~\cite{McIver2020}. This optical framework enables robust contactless detection: the ultrafast activation of the AHE and magnetization $\mathcal{M}$ generates time-varying currents and magnetizations ($\partial \mathcal{M}/\partial t$) that act as easily isolated terahertz emission sources. Additionally, because the multi-color drive induces macroscopic magnetizations rather than purely edge-localized spin accumulations, time-resolved magneto-optic Kerr effect (TR-MOKE) spectroscopy~\cite{PhysRevLett.119.087203} can directly probe the center of the pumped spot, circumventing boundary diffraction limits. Nonlocal spin transport and Floquet ARPES can further resolve anisotropic responses. Finally, off-resonant pumped driving at 1.5–3 V/nm suppresses heating; once the pulse passes, the virtual-photon-sustained Floquet state collapses, and residual magnetizations decay over the material's intrinsic femtosecond-to-picosecond spin relaxation time. 
		
		\blue{\textit{Conclusion}}---In this work, we developed a Floquet framework for controlling charge and spin transport in 2D $d$-wave altermagnets (readily extendable to $g$- and $i$-wave orders) with an out-of-plane Néel vector and Rashba spin--orbit coupling. Combining Floquet--Magnus theory with semiclassical Boltzmann transport and the Kubo--Berry curvature formalism, we showed how periodic driving enables symmetry engineering of transport beyond equilibrium constraints. Monochromatic driving preserves the residual $C_{2v}$ symmetry of the altermagnetic state, yielding strong longitudinal anisotropies, finite charge AMR and AHE, and light polarization-controlled spin-current reversal, while transverse responses remain symmetry-forbidden. In contrast, bichromatic $\omega$--$2\omega$ driving reduces the symmetry toward $C_1$ through harmonic-sector interference, generating both in-plane and out-of-plane magnetizations and activating transverse charge and spin transport, including finite T-AMR. The response is tunable through light polarization, phase, and chirality, with opposite circular-field chiralities producing the strongest signals. Comparable magnetizations would require effective magnetic fields of hundreds of tesla, whereas Floquet engineering achieves them without static fields. These results establish multi-frequency Floquet driving as a versatile route for symmetry-controlled transport in unconventional altermagnets.
		
		\blue{\textit{Acknowledgments}}---M.\,Y. and J.\,K.\,F. were supported by the Department of Energy, Office of Basic Energy Sciences, Division of Materials Sciences and Engineering under Contract No. DE-FG02-08ER46542 for the formal developments, the analytical/numerical work, and writing of the manuscript. 
        J.\,K.\,F. was also supported by the McDevitt bequest at Georgetown University. M.\,Y. acknowledges the hospitality of Uppsala University during his visit, where part of this work was performed. J.\,S. was supported by the DOE-Office of Science, DE-SC0026038, for the analysis of the results and writing of the manuscript. P.\,M.\,G is funded by the European Union through a MSCA Postdoctoral Fellowship (Project No. 101145915) for writing the manuscript.
        
        \blue{\textit{Data availability}}---The data that support the findings of this article are openly available at~\cite{Zenodo}.
        
	}
    
	\bibliography{bib.bib}
    
	{\allowdisplaybreaks
		\onecolumngrid
		\subsection{\large End Matter}
		\twocolumngrid
		\hypertarget{mylinkA}{\blue{\textit{Effective Floquet Hamiltonian}}}---In the off-resonant regime, following the detailed derivation provided in the SM~\cite{SM}, the effective Hamiltonian retains the form{\small\begin{equation}
				\mathcal{H}_{\rm eff}(\mathbf{k}) =
				h_0^{\rm pol}(\mathbf{k})\sigma_0 + h_x^{\rm pol}(\mathbf{k})\sigma_x
				+ h_y^{\rm pol}(\mathbf{k})\sigma_y + h_z^{\rm pol}(\mathbf{k})\sigma_z.
		\end{equation}}
		
		For linearly polarized bichromatic drive, the components are{\small\begin{subequations}\label{eq_13}
				\begin{align}
					h_0^{\rm lin}(\mathbf{k}) &= \frac{\hbar^2}{2m_e}(k_x^2+k_y^2)
					+ \frac{e^2 \mathcal{A}^2}{4m_e}(1+\mathcal{S}^2),\\[4pt]
					h_x^{\rm lin}(\mathbf{k}) &= -\lambda k_y
					+ \frac{e^3\lambda\mathcal{A}^3 \mathcal{S}\sin\phi}{8\hbar^2 m_e\omega}
					\left(3M_{1g}\cos\eta + 4M_{2g}\sin\eta\right),\label{eq_13b}\\[4pt]
					h_y^{\rm lin}(\mathbf{k}) &= \lambda k_x
					- \frac{e^3\lambda\mathcal{A}^3 \mathcal{S}\sin\phi}{8\hbar^2 m_e\omega}
					M_{1g}\sin\eta,\label{eq_13c}\\[4pt]
					h_z^{\rm lin}(\mathbf{k}) &= M_{\mathbf{k}}^{d}
					+ \frac{e^2\mathcal{A}^2}{4m_e}
					\left[
					M_{1g}
					+ \mathcal{S}^2\left(M_{1g}\cos2\eta + M_{2g}\sin2\eta\right)
					\right].\label{eq_13d}
				\end{align}
		\end{subequations}}
		For bicircular drive, we obtain{\small\begin{subequations}\label{eq_14}
				\begin{align}
					&h_0^{\rm cir}(\mathbf{k}) = \frac{\hbar^2}{2m_e}(k_x^2+k_y^2)
					+ \frac{e^2 \mathcal{A}^2}{4m_e}(1+\mathcal{S}^2),\\[4pt]
					&h_x^{\rm cir}(\mathbf{k}) = -\lambda k_y
					+ \frac{e^2\lambda\mathcal{A}^2(2c+\tilde{c}\mathcal{S}^2)}{2\hbar\omega m_e}
					(M_{1g}k_y - M_{2g}k_x) \notag\\ 
					&\quad + \frac{e^3\lambda\mathcal{A}^3\mathcal{S}}{4\hbar^2\omega m_e}
					\Big[(1-2c\tilde{c})M_{1g}\sin\phi + (c-2\tilde{c})M_{2g}\cos\phi\Big],\\[4pt]
					&h_y^{\rm cir}(\mathbf{k}) = \lambda k_x
					+ \frac{e^2\lambda\mathcal{A}^2(2c+\tilde{c}\mathcal{S}^2)}{2\hbar\omega m_e}
					(M_{1g}k_x + M_{2g}k_y) \notag\\
					&\quad + \frac{e^3\lambda\mathcal{A}^3\mathcal{S}}{4\hbar^2\omega m_e}
					\Big[(c\tilde{c}-2)M_{2g}\sin\phi + (2c-\tilde{c})M_{1g}\cos\phi\Big],\\[4pt]
					&h_z^{\rm cir}(\mathbf{k}) = M_{\mathbf{k}}^{d}
					- \frac{e^2\lambda^2\mathcal{A}^2(2c+\tilde{c}\mathcal{S}^2)}{2\hbar^3\omega}.\label{eq_14d}
				\end{align}
		\end{subequations}}
		And finally, for a hybrid circular–linear drive, we achieve{\small\begin{subequations}\label{eq_15}
				\begin{align}
					&h_0^{\rm hyb}(\mathbf{k}) = \frac{\hbar^2}{2m_e}(k_x^2+k_y^2)
					+ \frac{e^2 \mathcal{A}^2}{4m_e}(2+\mathcal{S}^2),\\[4pt]
					&h_x^{\rm hyb}(\mathbf{k}) = -\lambda k_y
					+ \frac{e^2\lambda\mathcal{A}^2}{4\hbar^2 m_e\omega}
					\Big[
					4c\hbar(M_{1g}k_y - M_{2g}k_x) \notag\\
					&\quad + \mathcal{A}\mathcal{S}e
					\big(
					M_{1g}\cos\eta\sin\phi + 2M_{2g}\sin\eta\sin\phi \notag\\
					&\quad + c(M_{2g}\cos\eta\cos\phi - 2M_{1g}\sin\eta\cos\phi)
					\big)
					\Big],\\[4pt]
					&h_y^{\rm hyb}(\mathbf{k}) = \lambda k_x
					+ \frac{e^2\lambda\mathcal{A}^2}{4\hbar^2 m_e\omega}
					\Big[
					4c\hbar(M_{1g}k_x + M_{2g}k_y) \notag\\
					&\quad + \mathcal{A}\mathcal{S}e
					\big(
					-2M_{2g}\cos\eta\sin\phi + M_{1g}\sin\eta\sin\phi \\
					&\quad + c(2M_{1g}\cos\eta\cos\phi + M_{2g}\sin\eta\cos\phi)
					\big)
					\Big],\\[4pt]
					&h_z^{\rm hyb}(\mathbf{k}) = M_{\mathbf{k}}^{d}
					+ \frac{e^2\mathcal{A}^2}{4m_e}
					\Big[
					- \frac{4c\lambda^2 m_e}{\hbar^3\omega}
					\notag \\ &\quad+ \mathcal{S}^2(M_{1g}\cos2\eta + M_{2g}\sin2\eta)
					\Big].
				\end{align}
		\end{subequations}}
		
		\hypertarget{mylinkB}{\blue{\textit{Conductivities in the pristine $d$-wave altermagnet}}}---To compute longitudinal conductivities in the pristine (no RSOC) $d$-wave AM, we evaluate~[$\varepsilon = \hbar^2/2 m_e$]\begin{equation}\label{eq_15n}
			I_x^\nu=\int \frac{d^2k}{(4\pi^2)}\,v_x^2\,\delta(\epsilon_\nu-\mu),
		\end{equation}with $\epsilon_\nu= \varepsilon(k_x^2+k_y^2)
		+\nu\varepsilon M_d (k_x^2-k_y^2)\cos(2\theta)) +2\nu\varepsilon M_d k_xk_y\sin(2\theta),$ and $v_x=\hbar^{-1}\partial_{k_x}\epsilon_\nu$. After rotation $\mathbf{k}'=R(\theta)\mathbf{k}$ and rescaling
		$\tilde{k}_x=\sqrt{\varepsilon+\nu\varepsilon M_d}\,k_x'$, 
		$\tilde{k}_y=\sqrt{\varepsilon-\nu\varepsilon M_d}\,k_y'$,
		the dispersion becomes $\epsilon=\tilde{k}^2$ with $\tilde{k}_F=\sqrt{\mu}$ and $d^2k=\frac{d^2\tilde{k}}{\sqrt{(\varepsilon+\nu\varepsilon M_d)(\varepsilon-\nu\varepsilon M_d)}}.$ The velocity reads
		\begin{equation}
			v_x\hbar=
			2\sqrt{\varepsilon+\nu\varepsilon M_d}\,\tilde{k}_x\cos\theta
			-2\sqrt{\varepsilon-\nu\varepsilon M_d}\,\tilde{k}_y\sin\theta.
		\end{equation}
		
		Using polar coordinates and $\int d^2\tilde{k}\,\delta(\tilde{k}^2-\mu)=\frac12\int d\phi$, we obtain $I_x^\nu=
		\frac{\mu}{2\pi\hbar^2}
		\frac{1+\nu M_d\cos(2\theta)}{\sqrt{1-M_d^2}}.$ Hence\begin{align}
			\sigma_{xx}^c = \frac{e^2\tau\mu}{\pi\hbar^2\sqrt{1-M_d^2}},\qquad
			\sigma_{xx}^s = -\frac{e\tau\mu M_d\cos(2\theta)}{4\pi\hbar\sqrt{1-M_d^2}}.
		\end{align}For $M_d=0$, the isotropic result is recovered: $ \sigma_{xx}^c= \frac{e^2 \tau \mu}{\pi\hbar^2}$, the inverse of resistance) and $\sigma_{xx}^s = 0$. The denominator indicates a divergence at $M_d = 1$, corresponding to a van Hove singularity associated with band flattening along one direction.\begin{figure}[b]
			\centering
			\includegraphics[width=0.95\linewidth]{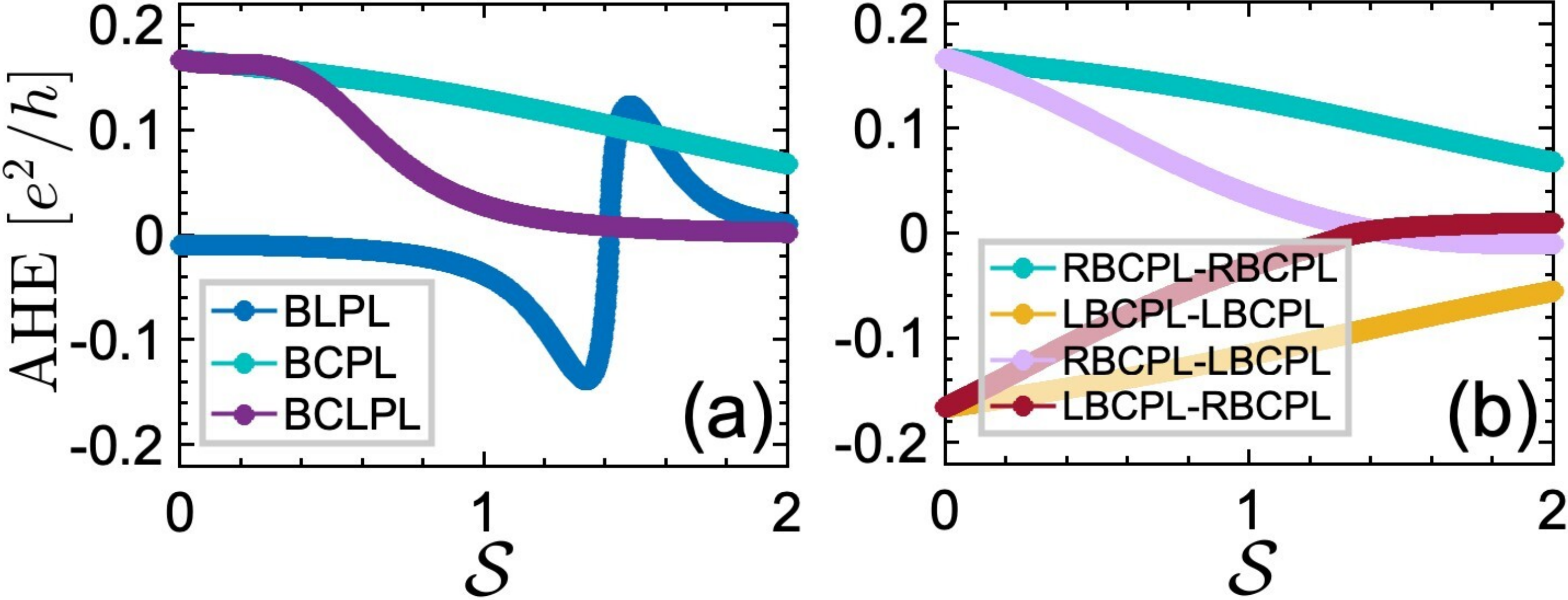}
			\caption{AHE of a $d_{x^2-y^2}$-wave AM with an out-of-plane Néel vector and RSOC versus $\mathcal{S}$ at fixed $e\mathcal{A}/\hbar k_F=0.7$. (a) BLPL, BCPL ($c=\tilde c=+1$), and BCLPL ($c=+1$). (b) RBCPL--RBCPL, LBCPL--LBCPL, RBCPL--LBCPL, and LBCPL--RBCPL configurations, highlighting the role of relative chirality. The BLPL response changes sign at $\mathcal{S}\approx\sqrt{2}$, while BCPL and BCLPL are progressively suppressed. A similar critical point appears for opposite circular chiralities. Parameters: $M_d=0.7$ and $\lambda=0.2$ eV\,\AA.}
			\label{f5}
		\end{figure} 
		
		Similarly, $I_y^\nu=
		\frac{\mu}{2\pi\hbar^2}
		\frac{1-\nu M_d\cos(2\theta)}{\sqrt{1-M_d^2}},$ giving\begin{align}
			\sigma_{yy}^c = \frac{e^2\tau\mu}{\pi\hbar^2\sqrt{1-M_d^2}},\qquad
			\sigma_{yy}^s = +\frac{e\tau\mu M_d\cos(2\theta)}{4\pi\hbar\sqrt{1-M_d^2}}.
		\end{align}For $M_d=0$, one recovers the isotropic result, i.e., $ \sigma_{yy}^c= \frac{e^2 \tau \mu}{\pi \hbar^2}$ and $\sigma_{yy}^s = 0$. For the transverse response, $I_{xy}^\nu=\frac{\mu}{2\pi}\frac{\nu M_d\sin(2\theta)}{\sqrt{1-M_d^2}},$ we get\begin{align}\label{eq_19n}
			\sigma_{xy}^c = 0,\qquad
			\sigma_{xy}^s = -\frac{e\tau\mu M_d\sin(2\theta)}{4\pi\hbar\sqrt{1-M_d^2}}.
		\end{align}For \(M_d = 0\), both charge and spin Hall conductivities vanish.
		
		\hypertarget{mylinkD}{\blue{\textit{Anomalous Hall effect under bichromatic driving}}}---The AHE is strongly tunable by the relative amplitude $\mathcal{S}$. As shown in Fig.~\ref{f5}, in the BLPL configuration, it evolves nonmonotonically, and changes sign near $\mathcal{S}\!\approx\!\sqrt{2}$. By contrast, BCPL ($c=\tilde c=+1$) and BCLPL ($c=+1$) exhibit a monotonic suppression without sign reversal. For circular drives, reversing both chiralities reverses the sign of the AHE, while opposite chiralities strongly suppress it through partial Berry-curvature cancellation, yielding a sign change near $\mathcal{S}\!\approx\!\sqrt{2}$. These results demonstrate optical control of the Floquet Hall response through both light polarization and relative amplitude.\begin{figure}[t]
			\centering
			\includegraphics[width=0.8\linewidth]{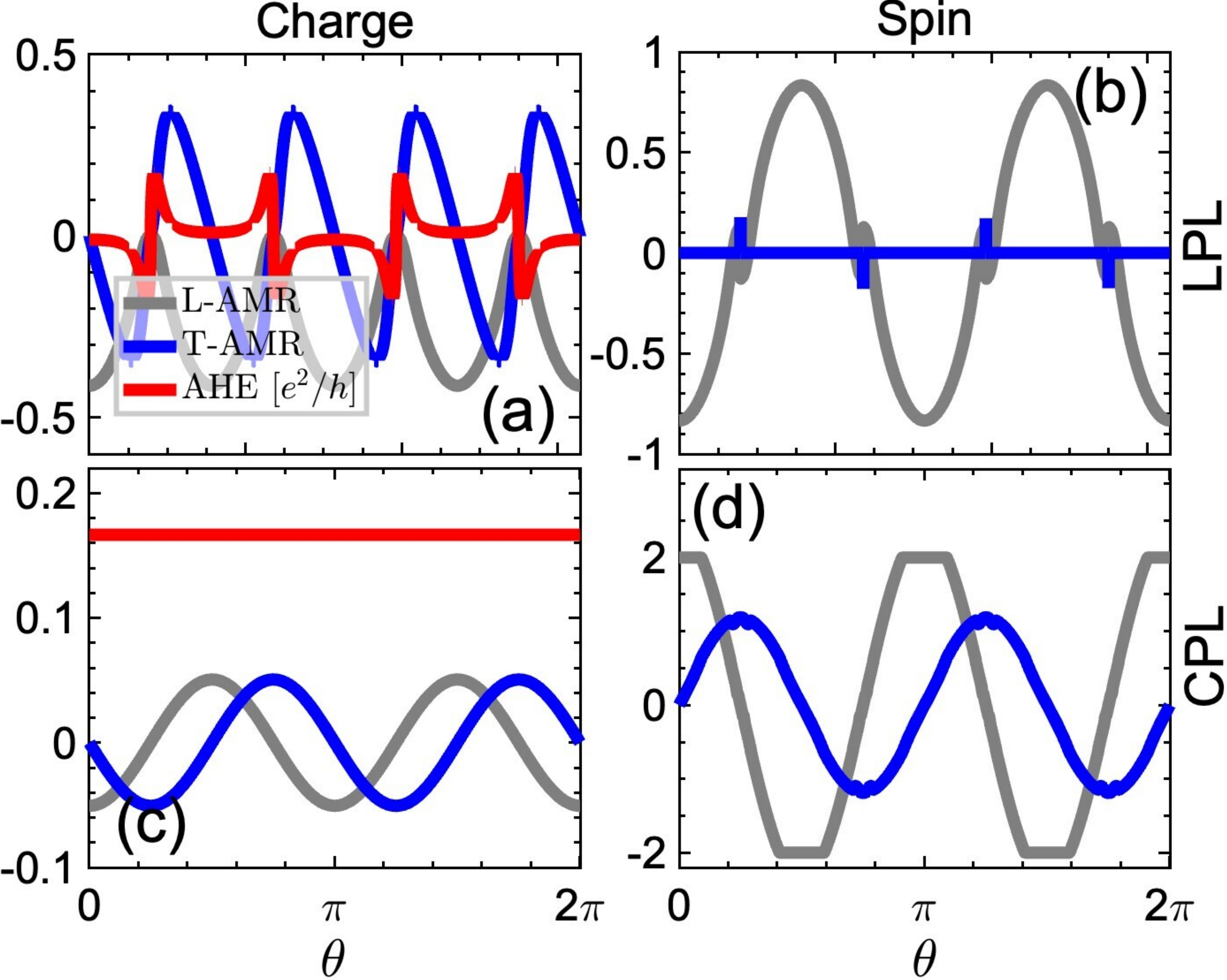}
			\caption{Charge (left column) and spin (right column) transport in a 2D $d$-wave AM with RSOC under monochromatic linear (a,b) and circular (c,d) driving as a function of the order-parameter orientation $\theta$ at $e\mathcal{A}/\hbar k_F=0.7$. Under LPL, $\theta$ strongly modulates both AMRs and the AHE, whereas under CPL it affects only the AMRs while leaving the AHE unchanged, offering a light polarization-resolved probe of altermagnetic order. Parameters: $M_d=0.7$, $\lambda=0.2$ eV\,\AA, and $c=+1$.}
			\label{f6}
		\end{figure}
		
		\hypertarget{mylinkD}{\blue{\textit{Effects of crystalline orientation of the order parameter}}}---We examine the dependence on the crystalline orientation $\theta$ of the order parameter under monochromatic driving [Fig.~\ref{f6}] at $e\mathcal{A}/\hbar k_F\simeq0.7$. Under both linear and circular driving, the charge L-AMR and AHE exhibit pronounced oscillations with sharp extrema at high-symmetry orientations. The charge T-AMR undergoes sign reversals with $\theta$, reflecting orientation-controlled transverse transport. The spin response shows larger L-AMR oscillations and small T-AMR, with sharper extrema than in the charge sector. 
        
        Under circular polarization [Fig.~\ref{f6}(c)], the AHE becomes independent of $\theta$, while the AMRs exhibit only weak angular oscillations, indicating that circular driving largely averages out the intrinsic $d$-wave anisotropy. In contrast, the spin response [Fig.~\ref{f6}(d)] retains pronounced angular modulation, with oscillatory AMRs reflecting the underlying altermagnetic symmetry. The finite-field contribution strongly enhances the signal relative to equilibrium, highlighting its nonlinear Floquet origin. 
	}
    
	\onecolumngrid
	\vskip40mm
	
	{\allowdisplaybreaks
		
		\begin{center}
			\textbf{\large \vskip0mm Supplemental Materials for ``Floquet-induced anisotropic magnetoresistance and anomalous Hall effect in 2D $d$-wave altermagnets with Rashba spin-orbit coupling''}
            \vskip3.5mm
			Mohsen Yarmohammadi,$^1$ Pieter M. Gunnink,$^2$ Jairo Sinova,$^{2,3}$ and James K. Freericks,$^1$\vskip1mm
			\small $^1$\textit{Department of Physics, Georgetown University, Washington DC 20057, USA}\\
            \small $^2$\textit{Institute of Physics, Johannes Gutenberg-University Mainz, Staudingerweg 7, Mainz 55128, Germany}\\
            \small $^3$\textit{Department of Physics, Texas A$\&$M University, College Station, Texas 77843-4242, USA}\\
            (Dated: \today)
		\end{center}
		\setcounter{equation}{0}
		\makeatletter

		\setcounter{equation}{0}
		\renewcommand{\theequation}{S\arabic{equation}}
		\setcounter{figure}{0}
		\renewcommand{\thefigure}{S\arabic{figure}}
		\setcounter{section}{0}
		\renewcommand{\thesection}{S\arabic{section}}
		\setcounter{table}{0}
		\renewcommand{\thetable}{S\arabic{table}}

        \section{S1. Effective Floquet Hamiltonian for a bichromatically driven 2D $d$-wave altermagnet with Rashba spin-orbit coupling}

        We consider a 2D $d$-wave AM driven by a two-color, phase-delayed optical field. Because the driving frequency $\omega$ is assumed to be much larger than the relevant energy scales of the unperturbed system ($\hbar\omega \gg W$, where $W$ is the bandwidth), we employ the high-frequency Floquet expansion. The time-periodic Hamiltonian $H(\mathbf{k}, t) = H(\mathbf{k}, t + 2\pi/\omega)$ can be expanded into its Fourier modes via
		\begin{equation}
			\mathcal{H}_m^{\rm F}(\mathbf{k}) = \frac{\omega}{2\pi} \int_{0}^{2\pi/\omega} H(\mathbf{k},t) e^{im\omega t} dt\,.
		\end{equation}
		To order $\mathcal{O}(\omega^{-1})$, the effective steady-state Floquet Hamiltonian is given by the time-average (zeroth-order) plus the dynamic commutator corrections (first-order):
		\begin{equation}
			\mathcal{H}^{\rm eff}(\mathbf{k}) = \mathcal{H}_0^{\rm F}(\mathbf{k}) + \sum_{m=1}^{\infty} \frac{1}{m\hbar\omega} \left[ \mathcal{H}_{-m}^{\rm F}(\mathbf{k}), \mathcal{H}_{m}^{\rm F}(\mathbf{k}) \right]\,.
		\end{equation}
		By isolating each Floquet mode into the Pauli basis $\mathcal{H}_m^{\rm F} = C_m\sigma_0 + \mathbf{V}_m \cdot \boldsymbol{\sigma}$, the commutator simplifies fundamentally via the Pauli algebra $[\sigma_i, \sigma_j] = 2i\epsilon_{ijk}\sigma_k$ as
		\begin{equation}
			\left[ \mathcal{H}_{-m}^{\rm F}, \mathcal{H}_{m}^{\rm F} \right] = \left[ \mathbf{V}_{-m} \cdot \boldsymbol{\sigma}, \mathbf{V}_m \cdot \boldsymbol{\sigma} \right] = 2i (\mathbf{V}_{-m} \times \mathbf{V}_m) \cdot \boldsymbol{\sigma}\,.
		\end{equation}
		Because the Hamiltonian is Hermitian, we obtain $\mathbf{V}_{-m} = \mathbf{V}_m^*$. Decomposing the vector into real and imaginary parts $\mathbf{V}_m = \mathbf{V}_R + i\mathbf{V}_I$ yields
		\begin{equation}
			\mathbf{V}_m^* \times \mathbf{V}_m = (\mathbf{V}_R - i\mathbf{V}_I) \times (\mathbf{V}_R + i\mathbf{V}_I) = 2i (\mathbf{V}_R \times \mathbf{V}_I)\,.
		\end{equation}
		
		The unperturbed $d$-wave AM Hamiltonian is
		\begin{equation}
			H_d(\mathbf{k}) = \varepsilon(k_x^2 + k_y^2)\sigma_0 + \lambda(k_x\sigma_y-k_y\sigma_x) + \varepsilon \big[M_{1g}(k_x^2-k_y^2) + 2M_{2g} k_xk_y\big]\sigma_z\,,
		\end{equation}
		where $\varepsilon=\hbar^2/(2m_e)$ defines the inverse effective mass, $\lambda$ is the RSOC, and $M_{1,2}$ define the altermagnetic order parameters. Coupling to the radiation field is introduced via the Peierls substitution $\mathbf{k} \rightarrow  \mathbf{k} + \frac{e}{\hbar}\mathbf{A}(t)$. Expanding the quadratic momentum terms strictly separates the Hamiltonian by powers of the vector potential. The full time-dependent Hamiltonian $H(\mathbf{k}, t) = H_0(\mathbf{k}) + H_1(t) + H_2(t)$ takes the exact form
		\begin{subequations}\begin{align}
			H_1(t) =& \left[ \frac{2e\varepsilon}{\hbar}\big(k_x A_x + k_y A_y\big) \right]\sigma_0 + \frac{e\lambda}{\hbar}\big(A_x\sigma_y - A_y\sigma_x\big) + \frac{2e\varepsilon}{\hbar} \Big[ M_{1g}\big(k_x A_x - k_y A_y\big) + M_{2g}\big(k_x A_y + k_y A_x\big) \Big]\sigma_z\,, \\[10pt]
			H_2(t) =& \left[ \frac{e^2\varepsilon}{\hbar^2}\big(A_x^2 + A_y^2\big) \right]\sigma_0 + \frac{e^2\varepsilon}{\hbar^2} \Big[ M_{1g}\big(A_x^2 - A_y^2\big) + 2M_{2g} A_x A_y \Big]\sigma_z\,.
		\end{align}	\end{subequations}
		The two-color vector potential components are given by
		\begin{subequations}
			\begin{align}
				A_x(t) &= \mathcal{A}\cos(\omega t) + \mathcal{A}\mathcal{S}\cos\eta\cos(n\omega t+\phi)\,, \\
				A_y(t) &= \mathcal{A}c\sin(\omega t) + \mathcal{A}\tilde{c}\mathcal{S}\cos(\eta-\tilde{\eta})\cos(n\omega t+\phi - \tilde{\phi})\,.
			\end{align}
		\end{subequations}
		To condense the algebra, we define the harmonic amplitudes and phases as
		\begin{equation}
			S_x = \mathcal{A}\mathcal{S}\cos\eta, \quad S_y = \mathcal{A}\tilde{c}\mathcal{S}\cos(\eta-\tilde{\eta}), \quad \phi_x = \phi, \quad \phi_y = \phi - \tilde{\phi}
		\end{equation}
		Using the complex exponential definitions $\cos(\theta) = \frac{1}{2}(e^{i\theta} + e^{-i\theta})$ and $\sin(\theta) = \frac{i}{2}(e^{-i\theta} - e^{i\theta})$, we extract the exact Fourier coefficients for the fundamental ($m=1$) and $n$-th harmonic ($m=n$) by reading the coefficients of $e^{-im\omega t}$, i.e.,
		\begin{subequations}\begin{align}
			A_{x,1} &= \frac{\mathcal{A}}{2}, & A_{y,1} &= i\frac{\mathcal{A}c}{2}, \\
			A_{x,n} &= \frac{S_x}{2}e^{-i\phi_x}, & A_{y,n} &= \frac{S_y}{2}e^{-i\phi_y}\,.
		\end{align}\end{subequations}
		The crucial positive imaginary unit in $A_{y,1}$ dictates the circular polarization chirality.
		
		The $H_2(t)$ sector contains $Q_1(t) = A_x^2 - A_y^2$ and $Q_2(t) = 2A_x A_y$. If the harmonic order is precisely $n=2$, the fundamental wave beating against itself generates new static $2\omega$ fields. For example, expanding $A_x^2$ leads to{\small\begin{align}
			A_x^2(t) &= \frac{\mathcal{A}^2}{4}\left( e^{i2\omega t} + e^{-i2\omega t} + 2 \right) + \frac{S_x^2}{4}\left( e^{i4\omega t} + e^{-i4\omega t} + 2 \right) + \frac{\mathcal{A}S_x}{2} \left( e^{i(\omega t + \phi_x)} + e^{-i(\omega t + \phi_x)} + e^{i(3\omega t + \phi_x)} + e^{-i(3\omega t + \phi_x)} \right)\,.
		\end{align}}
		Extracting the $m=1$ ($e^{-i\omega t}$) and $m=2$ ($e^{-i2\omega t}$) modes from all such pairs yields:
		\begin{subequations}\begin{align}
			(Q_1)_1 &= \Big( \frac{\mathcal{A}S_x}{2}e^{-i\phi_x} + i\frac{\mathcal{A}c S_y}{2}e^{-i\phi_y} \Big)\delta_{n,2}\,, \\
			(Q_2)_1 &= \Big( \frac{\mathcal{A}S_y}{2}e^{-i\phi_y} - i\frac{\mathcal{A}c S_x}{2}e^{-i\phi_x} \Big)\delta_{n,2}\,, \\
			(Q_1)_2 &= \frac{\mathcal{A}^2}{4}(1 + c^2)\delta_{n,2}\,, \\
			(Q_2)_2 &= i\frac{\mathcal{A}^2 c}{2}\delta_{n,2}\,.
		\end{align}\end{subequations}
		
		The time-averaged Hamiltonian $\mathcal{H}_0^{\rm F} = H_0 + \langle H_1 \rangle + \langle H_2 \rangle$ characterizes the static AC Stark shifts. Since $\mathbf{A}(t)$ has no strictly DC components, $\langle H_1 \rangle = 0$. The shifts emerge purely from $\langle H_2 \rangle$:
			\begin{subequations}\begin{align}
			\langle A_x^2 \rangle &= \frac{\mathcal{A}^2 + S_x^2}{2}, \quad \langle A_y^2 \rangle = \frac{\mathcal{A}^2c^2 + S_y^2}{2}\,, \\
			\langle 2 A_x A_y \rangle &= S_x S_y \langle \cos(n\omega t + \phi_x) \cos(n\omega t + \phi_y) \rangle = S_x S_y \cos(\phi_x - \phi_y) = S_x S_y \cos\tilde{\phi}\,.
		\end{align}\end{subequations}
		This yields
		\begin{equation}
			\mathcal{H}_0^{\rm F}(\mathbf{k}) = \big(\varepsilon k^2 + \Delta h_0^{(0)}\big)\sigma_0 + \lambda(k_x\sigma_y-k_y\sigma_x) + \big(\varepsilon M_{\mathbf{k}}^d + \Delta h_z^{(0)}\big)\sigma_z\,,
		\end{equation}
		with shifts
		\begin{subequations}\begin{align}
			\Delta h_0^{(0)} &= \frac{e^2\varepsilon}{2\hbar^2} \Big[ \mathcal{A}^2(1 + c^2) + S_x^2 + S_y^2 \Big]\,, \\
			\Delta h_z^{(0)} &= \frac{e^2\varepsilon}{2\hbar^2} \Big[ M_{1g}\big(\mathcal{A}^2(1 - c^2) + S_x^2 - S_y^2\big) + 2M_{2g} S_x S_y \cos\tilde{\phi} \Big]\,.
		\end{align}\end{subequations}
		
		Using the identity $2i(\mathbf{V}_R \times \mathbf{V}_I)$, we extract the topological field corrections. The Pauli vector $\mathbf{V}_1 = \mathbf{V}_{1,R} + i\mathbf{V}_{1,I}$ from $H_1$ is
		\begin{subequations}\begin{align}
			\mathbf{V}_{1,R} &= \left( 0, \frac{e\lambda\mathcal{A}}{2\hbar}, \frac{e\varepsilon\mathcal{A}}{\hbar}(M_{1g} k_x + M_{2g} k_y) \right) \,,\\
			\mathbf{V}_{1,I} &= \left( -\frac{e\lambda\mathcal{A}c}{2\hbar}, 0, \frac{e\varepsilon\mathcal{A}c}{\hbar}(-M_{1g} k_y + M_{2g} k_x) \right)\,.
		\end{align}\end{subequations}
		Taking the cross product $\mathbf{V}_{1,R} \times \mathbf{V}_{1,I}$ and multiplying by the prefactor $-\frac{2}{\hbar\omega}$ evaluates to:
		\begin{subequations}\begin{align}
			\Delta h_x^{(k, 1)} &= \frac{2e^2\lambda\varepsilon\mathcal{A}^2 c}{\hbar^3\omega} (M_{1g} k_y - M_{2g} k_x) \,,\\
			\Delta h_y^{(k, 1)} &= \frac{2e^2\lambda\varepsilon\mathcal{A}^2 c}{\hbar^3\omega} (M_{1g} k_x + M_{2g} k_y)\,, \\
			\Delta h_z^{(1)} &= -\frac{e^2\lambda^2\mathcal{A}^2c}{\hbar^3\omega}\,.
		\end{align}\end{subequations}
		
		The harmonic Pauli vector requires tracking the arbitrary phases $\phi_x, \phi_y$, i.e.,
			\begin{subequations}\begin{align}
			V_{x,n} &= -\frac{e\lambda S_y}{2\hbar}e^{-i\phi_y}, \quad V_{y,n} = \frac{e\lambda S_x}{2\hbar}e^{-i\phi_x}\,,\\
			V_{z,n}^{(k)} &= \frac{e\varepsilon}{\hbar} \Big[ M_{1g}(k_x S_x e^{-i\phi_x} - k_y S_y e^{-i\phi_y}) + M_{2g}(k_x S_y e^{-i\phi_y} + k_y S_x e^{-i\phi_x}) \Big]\,.
		\end{align}\end{subequations}
		Evaluating the cross product $(\mathbf{V}_n^* \times \mathbf{V}_n)_z = V_{x,n}^* V_{y,n} - V_{y,n}^* V_{x,n}$ gives
		\begin{align}
			(\mathbf{V}_n^* \times \mathbf{V}_n)_z &= -\frac{e^2\lambda^2 S_x S_y}{4\hbar^2} \left( e^{-i(\phi_x - \phi_y)} - e^{i(\phi_x - \phi_y)} \right) = i\frac{e^2\lambda^2 S_x S_y}{2\hbar^2}\sin\tilde{\phi}\,.
		\end{align}
		Multiplying by $\frac{2i}{n\hbar\omega}$ generates the new dynamic shifts:	\begin{subequations}\begin{align}
			\Delta h_x^{(k, n)} &= \frac{2e^2\lambda\varepsilon S_x S_y}{n\hbar^3\omega} \sin\tilde{\phi}\; (M_{1g} k_y - M_{2g} k_x) \,,\\
			\Delta h_y^{(k, n)} &= \frac{2e^2\lambda\varepsilon S_x S_y}{n\hbar^3\omega} \sin\tilde{\phi}\; (M_{1g} k_x + M_{2g} k_y) \,,\\
			\Delta h_z^{(n)} &= -\frac{e^2\lambda^2 S_x S_y}{n\hbar^3\omega} \sin\tilde{\phi}\,.
		\end{align}\end{subequations}
		
		We neatly sum the $m=1$ and $m=n$ geometric symmetries by defining an effective chirality $c_{\rm eff}$
		\begin{equation}
			c_{\rm eff} = c + \frac{S_x S_y}{n\mathcal{A}^2}\sin\tilde{\phi} = c + \frac{\mathcal{S}^2}{n}\tilde{c}\cos\eta\cos(\eta-\tilde{\eta})\sin\tilde{\phi}\,.
		\end{equation}
		Setting $\alpha = \frac{4e^2\lambda\varepsilon\mathcal{A}^2}{\hbar^3\omega}$, the unified dynamic corrections are:
	\begin{subequations}
	\begin{align}
			\Delta h_x^{(k)} &= \alpha c_{\rm eff} (M_{1g} k_y - M_{2g} k_x)\,, \\
			\Delta h_y^{(k)} &= \alpha c_{\rm eff} (M_{1g} k_x + M_{2g} k_y)\,, \\
			\Delta h_z^{\rm dyn} &= -\frac{2e^2\lambda^2\mathcal{A}^2}{\hbar^3\omega} c_{\rm eff}\,.
		\end{align}\end{subequations}
	If $n=2$, the beating components from $H_2$ inject static terms into $V_{z,m}$. For $m=1$, $V_{z,1}^{\text{extra}} = R_1 + i I_1$ with
			\begin{subequations}\begin{align}
			R_1 &= \frac{e^2\varepsilon\mathcal{A}}{2\hbar^2} \Big[ M_{1g}(S_x\cos\phi_x + c S_y\sin\phi_y) + M_{2g}(S_y\cos\phi_y - c S_x\sin\phi_x) \Big] \,,\\
			I_1 &= \frac{e^2\varepsilon\mathcal{A}}{2\hbar^2} \Big[ M_{1g}(c S_y\cos\phi_y - S_x\sin\phi_x) - M_{2g}(c S_x\cos\phi_x + S_y\sin\phi_y) \Big]\,.
		\end{align}\end{subequations}
		For $m=2$, $V_{z,2}^{\text{extra}} = R_2 + i I_2$, we obtain
		\begin{equation}
			R_2 = \frac{e^2\varepsilon\mathcal{A}^2}{4\hbar^2} M_{1g}(1+c^2), \quad I_2 = \frac{e^2\varepsilon\mathcal{A}^2}{4\hbar^2} M_{2g}(2c)\,.
		\end{equation}
		Evaluating the commutators of these fields against the base Rashba Pauli vectors $\mathbf{V}_{1,2}$ yields precisely
		\begin{subequations}\begin{align}
			\Delta h_x^{\rm indep} =& \;\Gamma\delta_{n,2} \Big\{ M_{1g}\Big[\frac{3-c^2}{2}\cos\eta\sin\phi_x - 2c\tilde{c}\cos(\eta-\tilde{\eta})\cos\phi_y\Big] \nonumber \\
			&+ M_{2g}\Big[c\cos\eta\cos\phi_x + 2\tilde{c}\cos(\eta-\tilde{\eta})\sin\phi_y\Big] \Big\}\,, \\[10pt]
			\Delta h_y^{\rm indep} =& \;\Gamma\delta_{n,2} \Big\{ M_{1g}\Big[2c\cos\eta\cos\phi_x + \frac{3c^2-1}{2}\tilde{c}\cos(\eta-\tilde{\eta})\sin\phi_y\Big] \nonumber \\
			&+ M_{2g}\Big[c\tilde{c}\cos(\eta-\tilde{\eta})\cos\phi_y - 2c^2\cos\eta\sin\phi_x\Big] \Big\}\,.
		\end{align}\end{subequations}
		where $\Gamma = \frac{e^3\lambda\varepsilon\mathcal{A}^3\mathcal{S}}{\hbar^4\omega}$ measures the nonlinear interaction strength.
		
		The complete effective Hamiltonian is given by $\mathcal{H}^{\rm eff}(\mathbf{k}) = h_0 \sigma_0 + h_x \sigma_x + h_y \sigma_y + h_z \sigma_z$ with
		\begin{align}
			h_0 &= \varepsilon(k_x^2 + k_y^2) + \Delta h_0^{(0)}\,, \\[10pt]
			h_x &= k_y\Big[-\lambda + \alpha c_{\rm eff} M_{1g}\Big] - k_x\Big[\alpha c_{\rm eff} M_{2g}\Big] + \Delta h_x^{\rm indep}\,, \\[10pt]
			h_y &= k_x\Big[\lambda + \alpha c_{\rm eff} M_{1g}\Big] + k_y\Big[\alpha c_{\rm eff} M_{2g}\Big] + \Delta h_y^{\rm indep}\,, \\[10pt]
			h_z &= \varepsilon\Big[M_{1g}(k_x^2-k_y^2) + 2M_{2g}k_xk_y\Big] + \Delta h_z^{(0)} - \frac{2e^2\lambda^2\mathcal{A}^2}{\hbar^3\omega} c_{\rm eff}\,.
		\end{align}
		
		The bare RSOC parameters $(-\lambda k_y, \lambda k_x)$ are dynamically renormalized by terms proportional to $\alpha c_{\rm eff}$. Notably, this light-induced SOC acts purely as an \textit{anisotropic} extension strictly tied to the altermagnetic order ($M_{1g}, M_{2g}$), implying optical control over the symmetry of the spin-momentum locking. The out-of-plane Dirac mass $\Delta h_z^{\rm dyn} \propto c_{\rm eff}$ is responsible for breaking time-reversal symmetry globally. Remarkably, $c_{\rm eff}$ demonstrates that one can generate this gap \textit{even with strictly linearly polarized constituents} ($c=0$), provided a relative optical phase delay ($\tilde{\phi} \neq 0$) establishes an effective optical helicity in the two-color superposition. The emergence of $\Delta h_{x,y}^{\rm indep}$ purely for $n=2$ indicates an explicit breaking of spatial symmetries driven by the commensurability of the $1\omega \pm 2\omega$ beating. These constant, optically-induced in-plane Zeeman fields act as synthetic magnetic fields capable of driving steady-state orbital magnetization or domain-wall motion without bulk $k$-dependence.
            }
	
\end{document}